\documentclass[useAMS, usenatbib]{mn2e}

\usepackage{epsfig}

\usepackage{amssymb}

\usepackage{multirow}

\usepackage{enumitem}

\usepackage{ulem}

\usepackage{rotating}
\usepackage{natbib}
\usepackage{latexsym}
\bibpunct{(}{)}{;}{a}{}{,}

\usepackage[hyphens]{url}

\begin{document}

\newcommand{\ledd}{%
$L_\mathrm{Edd}$}

\newcommand{\fgl}{3FGL~J1306.8--4031}

\newcommand{\Msun}{M$_\mathrm{\odot}$}

\def\rem#1{#1}
\def\hide#1{}

\def \aj {AJ}
\def \mnras {MNRAS}
\def \apj {ApJ}
\def \apjs {ApJS}
\def \apjl {ApJL}
\def \aap {A\&A}
\def \aapr {A\&ARv}
\def \nat {Nature}
\def \araa {ARAA}
\def \pasp {PASP}
\def \aaps {AAPS}
\def \prd {PhRvD}
\def \apss {ApSS}

\newcommand{\specialcell}[2][c]{%
  \begin{tabular}[#1]{@{}c@{}}#2\end{tabular}}

\author[M. Linares]{
\parbox[t]{\textwidth}{
\raggedright 
Manuel Linares$^{1}$\thanks{manuel.linares@upc.edu},
}
\vspace*{4pt}\\
$^1$ Departament de F{\'i}sica, EEBE, Universitat Polit{\`e}cnica de Catalunya, c/ Eduard Maristany 10, 08019 Barcelona, Spain
\\
}

 \title[A new redback in town]{The 26.3-hr orbit and multi-wavelength
   properties of the ``redback'' millisecond pulsar PSR~J1306--40}

\maketitle{}

%\vspace{-1cm}

\begin{abstract}

  We present the discovery of the variable optical and X-ray counterparts
  to the radio millisecond pulsar (MSP) PSR~J1306--40, recently
  discovered by \citeauthor{Keane17}
 We find that both the optical and X-ray fluxes are modulated with the
 same period, which allows us to measure for the first time the
 orbital period P$_{\rm orb}$=1.09716[6]~d.
The optical properties are consistent with a main sequence companion
with spectral type G to mid K and, together with the X-ray luminosity
(8.8$\times$10$^{31}$~erg~s$^{-1}$ in the 0.5--10~keV band, for a
distance of 1.2~kpc), confirm the redback classification of this pulsar.
 Our results establish the binary nature of PSR~J1306--40, which has the
 longest P$_{\rm orb}$ among all known compact binary MSPs in the
 Galactic disk.
 We briefly discuss these findings in the context of irradiation and
intrabinary shock emission in compact binary MSPs.
  
\end{abstract}


\vspace{-0.5cm}

\section{Introduction}
\label{sec:intro}

New nearby (D$\lesssim$4~kpc) millisecond pulsars (MSPs) in compact binaries
(orbital periods P$_{\rm orb}$$\lesssim$1~d) are being discovered in a
number of different ways.
These include radio timing observations \citep{Hessels11,Ray12},
``blind'' pulsation searches \citep{Pletsch12} and optical studies
\citep{Romani11,Kong12,Salvetti15,Linares17} of Fermi-LAT unidentified
GeV sources.
Compact binary MSPs are often classified according to the mass of the
companion star into ``black widows'' (M$_2 \sim$0.01~M$_\odot$) and
``redbacks'' (M$_2 \sim$0.1~M$_\odot$; \citealt{Roberts11}).

A panchromatic approach to finding new MSPs is important, since the
pulsations often prove difficult to detect in the radio and gamma-ray
bands, mostly due to occultation and acceleration effects along
  the orbit, respectively.
While systems with P$_{\rm orb}$ close to a day have been especially
elusive so far, a redback candidate in a 21-hr orbit was recently
discovered by \citet{Linares17} and \citet{Li16}.
This emergent population of MSPs has potentially far-reaching
implications for many fields of astrophysics, ranging from binary
evolution \citep{Benvenuto15} and astroparticle physics
\citep{Venter15} to the maximum neutron star mass and the equation of
state at supra-nuclear densities \citep[see, e.g.,][and references
  therein]{Ozel16}.

The MSP PSR~J1306--40 was recently discovered as part of an ongoing
radio survey \citep[SUPERB,][]{Keane17}, with a spin period of 2.2~ms
and a dispersion measure indicating a distance of D$\approx$1.2~kpc.
Even if no orbital solution could be obtained, \citet{Keane17} suggest
that this is a redback-type binary MSP eclipsed during a large
fraction of the orbit.
In order to confirm or reject this hypothesis, we have studied the
multi-wavelength properties of PSR~J1306--40 from the infrared to the
gamma-ray bands.
We report our results hereafter, including the discovery of optical
and X-ray flux modulations which reveal a 26.3 hr orbital period.

\begin{figure*}
%\centering
  \begin{center}
  \resizebox{2.0\columnwidth}{!}{\rotatebox{-90}{\includegraphics[]{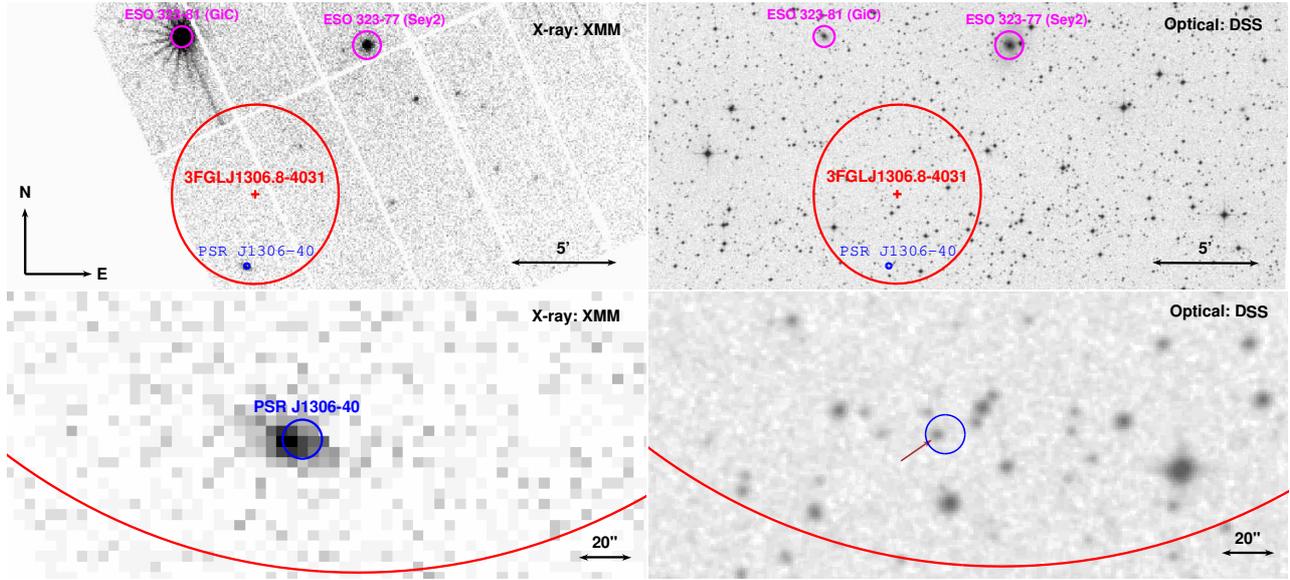}}}
%  \resizebox{1.0\columnwidth}{!}{\rotatebox{-90}{\includegraphics[]{f1b.eps}}}
%
  \caption{
{\it Left:} X-ray image of the field of \fgl\ (red ellipse) from the
longest XMM EPIC-PN observation (top) and zoom into the region of
PSR~J1306-40 (blue circle; bottom). Two nearby galaxies and bright
X-ray sources are marked (magenta circles).
{\it Right:} Optical DSS image of the field (top) and zoomed finding
chart (bottom), showing the radio location of PSR~J1306-40 (blue
circle) and the variable optical counterpart reported in this work
(SSS~J130656.3-403522; brown arrow).
} %
    \label{fig:chart}
%\epsscale{1.0}
 \end{center}
\end{figure*}

\section{Data Analysis and Results}
\label{sec:results}

\subsection{Optical}
\label{sec:phot}

We searched the Catalina Sky Survey catalog \citep[CSS,][]{Drake09},
and found one matching source (SSS~J130656.3-403522) at the radio
position of PSR~J1306--40 \citep[][]{Keane17}.
As we show in Figure~\ref{fig:chart} (right), this is the only optical
source within the radio error circle.
We find a refined optical position from the USNO-B1 catalog of
R.A.=13$^h$06$^m$56.30$^s$, DEC=-40$^\circ$35$'$23.3$''$ (J2000,
0.2$''$ astrometric uncertainty, \citealt{Monet03}; this is
  0.6$''$ from the CSS source position).
Thus our variable optical counterpart's position is fully consistent
with the 7''-radius radio location.
We downloaded the V band photometric CSS light curve, taken between
2005 and 2012 with the SSS 0.5-m telescope from nightly sequences of
typically four 30-s images.
In order to reduce the noise in the light curve, we filtered out data
taken at airmass$>$1.7 or with seeing worse than 7$''$, leaving a
total of 155 magnitude measurements.
We verified that relaxing these filters and using the full dataset
(with 257 data points) has no impact on the final period reported
below.
Furthermore, we inspected the CSS lightcurve rebinned into 50-d bins
and found no signs of a state change in the optical band (the
long-term averaged V band magnitude remains approximately constant).

\begin{figure}
%\centering
  \begin{center}
  \resizebox{1.0\columnwidth}{!}{\rotatebox{-90}{\includegraphics[]{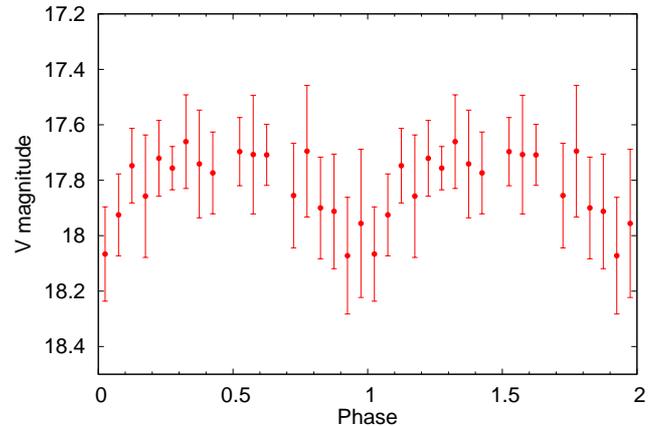}}}
  \caption{
V band light curve of the optical counterpart to PSR~J1306-40
(SSS~J130656.3-403522), folded at the orbital period P$_{\rm
  orb}$=1.09716[6]~d.
The epoch of inferior conjunction of the secondary star
(T0=56310.421$\pm$0.014~MJD, which defines phase=0) is estimated from
the X-ray light curve, which is modulated with the same period
(Sec.~\ref{sec:xray} for details).
Two cycles are plotted for clarity.
} %
    \label{fig:Vlc}
%\epsscale{1.0}
 \end{center}
\end{figure}

The CSS light curve shows clear signs of variability, with V
magnitudes generally in the 17.4--18.4 range.
We used a Lomb-Scargle periodogram to perform an initial search for
periodicities in the 0.1--10~d range, which includes all known redback
orbital periods.
Despite the aliasing due to nightly sampling and data gaps, we
identify the three strongest peaks at 1.10, 1.96 and 2.03~d.
Because the X-ray lightcurve of PSR~J1306--40 is modulated with the
same $\sim$1.10~d period (Section~\ref{sec:xray}), which shows the
highest power in the Lomb-Scargle periodogram, we focus the rest of
our analysis on this.
Applying phase dispersion minimization techniques
\citep{Stellingwerf78}, we obtain a refined measurement of the optical
photometric period P$_{\rm opt}$=1.09716[6]~d = 26.3319[14]~h.
We show in Figure~\ref{fig:Vlc} the corresponding optical light curve
folded at this period, which reveals one broad maximum and one
narrower minimum per cycle.
Some of the known redbacks show one maximum per orbital cycle due
to the strongly irratiated ``inner'' face of the companion star
\citep[e.g.,][]{Breton13}, while other redbacks feature two maxima
per orbit \citep[e.g.,][]{Li14}, due to the change in projected area
in the absence of strong irradiation (the so-called el.lipsoidal
modulation, see Section~\ref{sec:discussion} for further
discussion). We closely inspected the Lomb-Scargle periodogram around
2$\times$P$_{\rm opt}$, and found no periodicities consistent with
this value.

\subsection{X-ray}
\label{sec:xray}

The X-ray Multi-Mirror Mission ({\it XMM-Newton}) observed the field
of \fgl\ on 2006-02-07 (for 29~ksec in observation 0300240501) and
2013-01-17 (for 133~ksec in observation 0694170101), aiming at a
nearby Seyfert 2 galaxy (see Figure~\ref{fig:chart}, left).
This resulted in the serendipitous detection of a fainter X-ray
source, 3XMM~J130656.2-403523, which matches the radio position of
PSR~J1306-40 \citep[as pointed out by][see
  Fig.~\ref{fig:chart}]{Keane17}.
We analyzed the spectrum and light curve of the X-ray counterpart to
PSR~J1301-40 from the longest observation, using the pipeline
processing system products ({\sc SAS} version 20150701\_1240-14.0.0)
and the latest available response matrix (epn\_e3\_ff20\_sdY5\_v16.0).
The second half of the observation suffered from background flaring
periods, which had only minor effects on the background-corrected
light curve, but were excluded from the spectral analysis.

The average 0.2--12~keV EPIC-pn spectrum from the 2013 observation is
well fitted with an absorbed power law model (tbabs*power within {\sc
  Xspec}; \citealt{Arnaud96}) with absorbing column density N$_{\rm
  H}$=[6$\pm$1]$\times$10$^{20}$~cm$^{-2}$, a photon index
1.31$\pm$0.04 and an absorbed 0.5--10~keV flux of
[4.9$\pm$0.1]$\times$10$^{-13}$~erg~s$^{-1}$~cm$^{-2}$.
The corresponding 0.5--10~keV luminosity for the D=1.2~kpc distance
\citep{Keane17} is 8.8$\times$10$^{31}$~erg~s$^{-1}$, fully consistent
with that of redback MSPs in the radio pulsar state \citep[which
  also feature hard X-ray spectra with photon indices in the 1--2
  range;][]{Linares14c}.
The 2006 observation shows a similar photon index (1.3$\pm$0.1) and a
somewhat lower luminosity of 2.6$\times$10$^{31}$~erg~s$^{-1}$ (i.e.,
still consistent with the radio pulsar state).

\begin{figure}
%\centering
  \begin{center}
  \resizebox{1.0\columnwidth}{!}{\rotatebox{-90}{\includegraphics[]{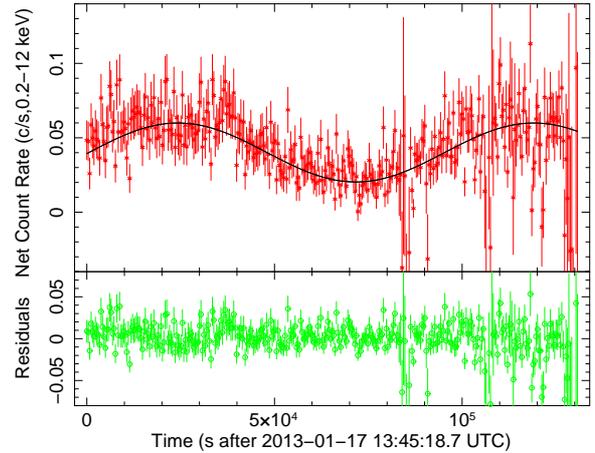}}}
  \caption{
XMM light curve of the X-ray counterpart to PSR~J1306-40 (top; net
0.2--12~keV count rate), modulated with the same period as the optical
light curve.
We also show our sinusoidal best-fit function (black line) and the
corresponding residuals (bottom).
 } %
    \label{fig:Xlc}
%\epsscale{1.0}
 \end{center}
\end{figure}

The background-corrected 0.2--12~keV light curve shows a clear
nearly-sinusoidal modulation (Figure~\ref{fig:Xlc}).
We fit the light curve with a simple sine model and find that the
X-ray (P$_{\rm X}$) and optical (P$_{\rm opt}$) periods are consistent
(within 2.7 sigma), while the fractional semi-amplitude is large,
close to 50\%.
In particular, a simple sine fit yields P$_{\rm X}$=1.14$\pm$0.03~d
and a reduced chi-squared of 1.31 for 342 degrees of freedom.
We also show in Figure~\ref{fig:Xlc} the results of fitting the X-ray
light curve with a sine function with the period fixed at P$_{\rm
  opt}$ (1.09716~d), which results into a similarly good fit (reduced
chi-squared of 1.31 for 343 degrees of freedom).

Interestingly, very similar X-ray orbital modulation is seen in
redback and black widow MSPs \citep{Bogdanov11,Gentile14}, with one
maximum and one minimum per orbital cycle in the X-ray lightcurve.
Therefore, we conclude that both X-ray and optical light curves of
PSR~J1306-40 are modulated at the orbital period P$_{\rm
  orb}$=1.09716[6]~d = P$_{\rm opt}$ = P$_{\rm X}$.
In those cases the minimum X-ray and optical fluxes correspond to the
same orbital phase (0 in our definition; 0.25 in the radio pulsar
definition), when the companion at inferior conjunction is thought to
partially occult the X-ray emmitting intra-binary shock between the
pulsar and the companion winds \citep{Bogdanov11}.
Using this analogy (i.e., assuming that the minimum X-ray flux
corresponds to phase=0) and our best fit to the XMM X-ray light curve,
we can constrain the epoch of inferior conjunction of the companion to
T0 = 56310.421$\pm$0.014~MJD (1 sigma statistical uncertainty).

\section{Discussion}
\label{sec:discussion}

We have presented the discovery of the optical and X-ray counterparts
to PSR J1306-40 \citep{Keane17}, which we show are both modulated with
the same period.
We thereby find the orbital period of this newly discovered MSP at
P$_{\rm orb}$=1.09716[6]~d, the longest among the currently known
compact binary MSPs in the Galactic disk.
The only known redback with longer P$_{\rm orb}$ is PSR J1740-5340, in
the globular cluster NGC~6397
\citep[P$_\mathrm{orb}$=1.35~d,][]{DAmico01}.
The rapid release of sky locations for new candidate or confirmed MSPs
allows multi-wavelength studies and a more efficient characterization
of this growing population.

We note that the orbital X-ray modulation that we have found in
PSR~J1306-40 (Sec.~\ref{sec:xray}), with an orbital period of about
26.3~hr, is very similar to that of the redback PSR~J1023+0038, with
P$_{\rm orb}$$\simeq$4.8~hr.
For a P$_{\rm orb}$$\sim$5 times longer (assuming similar total mass),
the orbital separation of PSR~J1306-40 is $\sim$3 times larger.
Our results show that in these conditions the shock responsible for
the X-ray emission can still be formed and observed.
As we discover MSPs with main sequence companions in wider orbits, we
will probe the pulsar wind and its interaction with that of the companion
star further away from the pulsar.
In particular, comparing shocks at different orbital separations may
tell us whether the companion wind is intrinsic/thermal or induced by
the pulsar \citep[see, e.g., discussion in][]{Harding90}.

Interestingly, we find one broad maximum and one narrow minimum per
orbital cycle in the optical light curve (Sec.~\ref{sec:phot},
Fig.~\ref{fig:Vlc}).
This is different than other redback MSPs with long orbital periods,
which tend to show two maxima per cycle \citep[see, e.g.,][and
  references therein]{Linares17}.
Thus we conclude that irradiation of the companion by the pulsar wind
is likely to play an important role in shaping the optical lightcurve
of PSR~J1306-40, despite its relatively wide orbit. This might be
  due to an energetic or beamed MSP wind, to a relatively faint
  companion or to a combination thereof.

We find no evidence of any optical or X-ray state change between 2005
and 2013, from the available XMM (2006, 2013) and CSS (2005--2012)
data sets.
Considering also the radio pulsar detections in June 2015 and
September 2016 \citep{Keane17}, and taking into account that accretion
disk states may last from months to years, this suggests that
PSR~J1306-40 has been in the radio pulsar (disk-free) state for at
least the last $\sim$11 years.

%\subsection{Infrared}
%\label{sec:IR}

Our variable X-ray and optical counterparts match a 2MASS source
\citep[2MASS~J13065627--4035233;][]{2mass} with near-infrared
magnitudes of J=16.26$\pm$0.12, H=15.99$\pm$0.17 and K=15.39$\pm$0.20.
% \citep{2mass,wise}
At longer wavelengths, we find a coincident WISE source
\citep[WISE~J130656.28-403523.3;][]{wise}, with magnitudes
w1=15.61$\pm$0.04, w2=15.77$\pm$0.11, w3$>$13.0 and w4$>$9.3.
The NOMAD/USNO-B1 catalog lists B, V, R and I magnitudes for this
object of $\simeq$18.4,17.7,18.1 and 17.5, respectively.
The reddening from infrared dust maps at the position of PSR J1306-40
\citep{Schlegel98} is E(B-V)=0.092$\pm$0.002~mag, which agrees with
the value inferred from our measured N$_{\rm H}$
(E(B-V)~$\sim$~N$_{\rm
  H}$/[3.1$\times$1.8$\times$10$^{21}$~cm$^{-2}$]~$\sim$~0.1~mag;
\citealt{Predehl95}).
Using this reddening, we estimate an intrinsic colour
(B-V)$\simeq$0.60, which is consistent with an early G star
\citep{Pecaut13}.

We find V$_{\rm CSS}$~$\simeq$~18~mag at minimum (Fig.~\ref{fig:Vlc};
V$\simeq$18.15~mag applying the CSS colour
correction\footnote{http://nesssi.cacr.caltech.edu/DataRelease/FAQ2.html\#improve}),
which corresponds to an extinction-corrected V$\simeq$17.85.
For the inferred D=1.2~kpc \citep{Keane17}, the absolute visual
magnitude is M$_V \sim$~7.45, close to that of a K5 main sequence star
\citep{Pecaut13}.
Therefore, while more accurate -- and phase-resolved-- optical
photometry and spectroscopic measurements are desirable and needed to
place tighter constraints on the binary parameters, we can estimate a
G-to-mid-K spectral type for the companion to PSR~J1306-40.
Finally, since all known black widows known are fainter than r$\sim$19~mag
\citep[e.g. PSR 1957+20, with absolute visual magnitude
  $\simeq$~10.5;][]{Djor88}, our optical counterpart to PSR~J1306-40
strongly favours a redback-type main sequence companion (unless the
distance measured by \citealt{Keane17} from the pulsar's dispersion measure
is severely overestimated).
The same is true for the X-ray luminosity that we find, close to
10$^{32}$~erg~s$^{-1}$ (Sec.~\ref{sec:xray}), which also points to a
redback MSP identification.

The Fermi-LAT source \fgl\ \citep{Acero15} includes PSR~J1306--40 in
its 4.3' error circle, and MSPs are well-known gamma-ray emitters.
However, while the flux variability (variability index 44) and
0.1-100~GeV luminosity ([2.5$\pm$0.2]$\times$10$^{33}$~erg~s$^{-1}$ at
D=1.2~kpc) are in line with the properties of virtually all
LAT-detected MSPs, the GeV spectrum is clearly at odds.
\fgl\ features a flat spectrum with little or no spectral curvature
(curvature significance 0.3), without any apparent cutoff up to at
least 10~GeV.
This is unlike any LAT spectrum of the known MSPs, with the
    possible exception of 3FGL~J1417.5-4402 \citep{Strader15,Camilo16}:
    a peculiar binary MSP with a giant companion in a 5.4-d orbit, possibly accreting, which also shows a flat spectrum in the LAT band.

There are no other catalogued LAT sources within at least two degrees
of \fgl.
However, the field contains two galaxies belonging to the same cluster,
only 8.8' and 8.3' from the center of the LAT position: ESO~323-77
(classified as Seyfert 2) and ESO~323-81, respectively.
Both galaxies are bright X-ray sources (see Figure~\ref{fig:chart},
top left panel), and active galaxies are known GeV emitters.
Since Fermi-LAT has an angular resolution of about 9' or worse, we
suggest that the LAT source is actually a blend of PSR~J1306-40 and
the nearby galaxies in the field.
That would attribute the flat spectrum to contamination from the
nearby galaxies, and it would also explain why \fgl\ is not centered
around PSR~J1306-40 but rather shifted towards the two galaxies
(Fig.~\ref{fig:chart}, top).
Detailed reanalysis of this LAT field may be able to address this
issue and resolve multiple sources.
That would in turn yield a more reliable measurement of the MSP
gamma-ray flux, and perhaps its orbital or spin modulations.\\

\textbf{Acknowledgments:}

We thank A. Harding for a discussion on MSP gamma-ray spectra, and
specifically for pointing out the possibility of source confusion in
the Fermi-LAT band, and J. Strader for pointing out the case of
3FGL~J1417.5-4402.
We thank G. Sala and A. Drake for guidance and suggestions on XMM and
CSS data products, respectively, P. Rodr{\'i}guez-Gil for useful
comments on an early draft, and the anonymous referee for a thorough
review which improved this manuscript.
This publication has made use of the SIMBAD database, operated at
CDS-Strasbourg-France, data obtained from the 3XMM XMM-Newton
serendipitous source catalogue compiled by the 10 institutes of the
XMM-Newton Survey Science Centre, as well as data products provided by
HEASARC, 2MASS and WISE.
The CSS survey is funded by the National Aeronautics and Space
Administration under Grant No. NNG05GF22G issued through the Science
Mission Directorate Near-Earth Objects Observations Program.
CRTS and CSDR2 are supported by the U.S.~National Science Foundation
under grant Ast-1413600.
Based on photographic data obtained using The UK Schmidt Telescope.
The Digitized Sky Survey was produced at the Space Telescope Science
Institute under US Government grant NAG W-2166.
M.L. is supported by EU's Horizon 2020 programme through a Marie
Sklodowska-Curie Fellowship (grant nr. 702638).

\vspace{0.2cm}

\noindent In memory of our esteemed colleague Enrique Garc{\'i}a-Berro.

\vspace{-0.5cm}

\bibliographystyle{mn2e}
%\bibliography{../biblio.bib}

\vspace{-0.6cm}

\end{document}